\documentclass[12pt]{article}
\pdfoutput=1
\usepackage{graphicx}
\usepackage{epsfig}
\usepackage{epstopdf}
\DeclareGraphicsExtensions{.pdf,.eps,.png,.jpg,.mps}
\usepackage{caption}
\usepackage{float}
\usepackage{color}

\usepackage{url}
\usepackage[implicit=false]{hyperref}
\usepackage{cite}

\def\lsim{\mathrel{\rlap{\lower3pt\hbox{\hskip0pt$\sim$}}
     \raise1pt\hbox{$<$}}}         
\def\gsim{\mathrel{\rlap{\lower4pt\hbox{\hskip1pt$\sim$}}
     \raise1pt\hbox{$>$}}}         

\usepackage{amsmath}
\usepackage{amsfonts}

\begin{document}
\begin{titlepage}

\centerline{\Large \bf Derivation of Spin Projection Operator}
\centerline{\Large \bf using Lagrange Interpolation Formula}
\medskip

\centerline{M.D. Zviadadze and Zura Kakushadze$^\S$$^\dag$\footnote{\, Zura Kakushadze, Ph.D., is the President of Quantigic$^\circledR$ Solutions LLC,
and a Full Professor at Free University of Tbilisi. Email: \href{mailto:zura@quantigic.com}{zura@quantigic.com}}}
\bigskip

\centerline{\em $^\S$ Quantigic$^\circledR$ Solutions LLC}
\centerline{\em 680 E Main St \#543, Stamford, CT 06901\,\,\footnote{\, DISCLAIMER: This address is used by the corresponding author for no
purpose other than to indicate his professional affiliation as is customary in
publications. In particular, the contents of this paper
are not intended as an investment, legal, tax or any other such advice,
and in no way represent views of Quantigic$^\circledR$ Solutions LLC,
the website \url{www.quantigic.com} or any of their other affiliates.
}}
\centerline{\em $^\dag$ Free University of Tbilisi, Business School \& School of Physics}
\centerline{\em 240, David Agmashenebeli Alley, Tbilisi, 0159, Georgia}
\medskip

\centerline{(November 30, 1988; in LaTeX form: January 18, 2020)\footnote{\, This note in Russian was published in 1989 in \cite{TSU}. This project was completed when I (ZK) was 13. For reasons outside my control, it was a few years before it was submitted to the journal. This English translation (by yours truly) closely follows the original Russian version, with minor changes such as equation formatting and some additional references.
}}

\bigskip
\medskip

\begin{abstract}
{}This note discusses how an operator analog of the Lagrange polynomial naturally arises in the quantum-mechanical problem of constructing an explicit form of the spin projection operator.
\end{abstract}
\medskip

\end{titlepage}

\newpage
{}The problem of interpolating a function $f(x)$ via a polynomial arises in many applications. One such interpolation formula has the form:
\begin{eqnarray}
 &&f(x) = P_N(x) + R_N(x)\\
 &&P_N(x) = \sum_{k=0}^N f(x_k)~\prod_{\ell=0;~\ell\neq k}^N ~{{x - x_\ell} \over {x_k - x_\ell}}\\
 &&R_N(x) = {f^{(N+1)}(\xi)\over(N+1)!}~\prod_{\ell=0}^N ~(x - x_\ell)\label{R}\\
 &&\min(x_k) < \xi < \max(x_k)
\end{eqnarray}
where $P_N(x)$ is the Lagrange polynomial, $R_N(x)$ is the residual term of the interpolation, and $x_k$ are the interpolation nodes \cite{IKS} (also see \cite{AS}). From (\ref{R}) it follows that $R(x) = 0$ and $f(x) = P_N(x)$ if $f(x)$ is a polynomial of order $n \leq N$.

{}This note discusses how an operator analog of the Lagrange polynomial naturally arises in the quantum-mechanical problem of constructing an explicit form of the projection operator ${\widehat P}_m({\widehat S}_z)$ of a general spin state onto the eigenstate $|m\rangle$ of the spin operator ${\widehat S}_z$ \cite{LL} (also see \cite{LL1}):
\begin{eqnarray}
 &&{\widehat S}_z|m\rangle = m|m\rangle\label{e1}\\
 &&m = -S, -S+1,\dots, S\\
 &&S = 0,{1\over 2},1,{3\over 2},\dots
\end{eqnarray}
The projection operator has the following properties:
\begin{eqnarray}
 &&{\widehat P}^2_m({\widehat S}_z) = {\widehat P}_m({\widehat S}_z)\label{e2}\\
 &&{\widehat P}_m({\widehat S}_z)|m^\prime\rangle = \delta_{mm^\prime}|m\rangle\label{e3}
\end{eqnarray}

{}Let us consider an arbitrary operator-valued function ${\widehat f}({\widehat S}_z)$. From (\ref{e1}), (\ref{e2}) and (\ref{e3}) it follows that
\begin{equation}
 {\widehat f}({\widehat S}_z) = \sum_{m=-S}^S f(m)~{\widehat P}_m({\widehat S}_z)
\end{equation}
On the other hand, we have the Maclaurin series:
\begin{equation}
 {\widehat f}({\widehat S}_z) = \sum_{n=0}^\infty {f^{(n)}(0)\over n!} ~{\widehat S}_z^n\label{series}
\end{equation}
From the properties of the operator ${\widehat S}_z$ it follows that
\begin{equation}
 \prod_{m = -S}^S ({\widehat S}_z - m) = 0
\end{equation}
Therefore, the powers ${\widehat S}_z^n$ with $n\geq 2S+1$ can be expressed as linear combinations of the powers ${\widehat S}_z^k$, where $k=0,1,2,\dots,2S$ for  half-integer values of $S$, and $k=1,2,\dots,2S$ for integer values of $S$. So, the infinite series (\ref{series}) at the end can be reduced to a polynomial of order $2S$. As a result, the interpolating Lagrange polynomial with the argument ${\widehat S}_z$, with the interpolation nodes chosen as the eigenvalues of the operator ${\widehat S}_z$, coincides with ${\widehat f}({\widehat S}_z)$. That is,
\begin{equation}
 {\widehat f}({\widehat S}_z) = \sum_{m=-S}^S (-1)^{S+m+\alpha_S}{f(m)\over (S-m)!(S+m)!} \prod_{n = -S;~n\neq m}^S ({\widehat S}_z - n) \label{eq.f}
\end{equation}
where $\alpha_S=1$ for half-integer values of $S$, and $\alpha_S=0$ for integer values of $S$. (In (\ref{eq.f}) the increments of $m$ and $n$ equal 1.)

{}Comparing (\ref{series}) and (\ref{eq.f}), we find an explicit form of the projection operator:
\begin{equation}
 {\widehat P}_m({\widehat S}_z) = {(-1)^{S+m+\alpha_S}\over (S-m)!(S+m)!} \prod_{n = -S;~ n\neq m}^S ({\widehat S}_z - n)\label{proj}
\end{equation}
Taking into account (\ref{e1}), one can readily verify that (\ref{proj}) satisfies (\ref{e2}).

{}Finally, let us list some examples:
\begin{eqnarray}
 &&S = {1\over 2}:~~~{\widehat P}_{\pm {1\over 2}}({\widehat S}_z) = {1\over 2} \pm {\widehat S}_z\\
 &&S = 1:~~~{\widehat P}_0({\widehat S}_z) = \left(1 + {\widehat S}_z\right)\left(1 - {\widehat S}_z\right)\\
 &&S = 1:~~~{\widehat P}_{\pm 1}({\widehat S}_z) = {1\over 2}~{\widehat S}_z\left({\widehat S}_z \pm 1\right)\\
 &&S = {3\over 2}:~~~{\widehat P}_{\pm {1\over 2}}({\widehat S}_z) = {1\over 2}\left({3\over 2} + {\widehat S}_z\right)\left({3\over 2} - {\widehat S}_z\right)\left({1\over 2} \pm {\widehat S}_z\right)\\
 &&S={3\over 2}:~~~{\widehat P}_{\pm {3\over 2}}({\widehat S}_z) = {1\over 6}\left({1\over 2} + {\widehat S}_z\right)\left({1\over 2} - {\widehat S}_z\right)\left({3\over 2} \pm {\widehat S}_z\right)
\end{eqnarray}

\end{document}